\documentclass[prl,twocolumn,showpacs,superscriptaddress,altaffilletter]{revtex4-1}

\usepackage{graphicx}  
\usepackage{dcolumn}   
\usepackage{bm}        
\usepackage{amssymb}   
\usepackage{epstopdf}
\usepackage{hyperref}
\usepackage{color}

\newcommand{\Dogus}{Dogus University, Istanbul, Turkey}
\newcommand{\Saclay}{IRFU, Centre d'\'Etudes Nucl\'eaires de Saclay (CEA-Saclay), Gif-sur-Yvette, France}
\newcommand{\CERN}{European Organization for Nuclear Research (CERN), Gen\`eve, Switzerland}
\newcommand{\INR}{Institute for Nuclear Research (INR), Russian Academy of Sciences, Moscow, Russia}
\newcommand{\MPE}{Max-Planck-Institut f\"{u}r Extraterrestrische Physik, Garching, Germany}
\newcommand{\Trieste}{Istituto Nazionale di Fisica Nucleare (INFN), Sezione di Trieste and Universit\`a di Trieste, Trieste, Italy}
\newcommand{\Zaragoza}{Grupo de Investigaci\'{o}n de F\'{\i}sica Nuclear y Astropart\'{\i}culas, Universidad de Zaragoza, Zaragoza, Spain }
\newcommand{\Chicago}{Enrico Fermi Institute and KICP, University of Chicago, Chicago, IL 60637, USA}
\newcommand{\Thessaloniki}{Aristotle University of Thessaloniki, Thessaloniki, Greece}
\newcommand{\Demokritos}{National Center for Scientific Research ``Demokritos'', Athens, Greece}

\newcommand{\Patras}{Physics Department, University of Patras, Patras, Greece}
\newcommand{\Athens}{National Technical University of Athens, Athens, Greece}

\newcommand{\Vancouver}{Department of Physics and Astronomy, University of British Columbia, Vancouver, Canada }
\newcommand{\Darmstadt}{Technische Universit\"{a}t Darmstadt, IKP, Darmstadt, Germany}
\newcommand{\Frankfurt}{Johann Wolfgang Goethe-Universit\"at, Institut f\"ur Angewandte Physik, Frankfurt am Main, Germany}
\newcommand{\Zagreb}{Rudjer Bo\v{s}kovi\'{c} Institute, Zagreb, Croatia}
\newcommand{\MPP}{Max-Planck-Institut f\"{u}r Physik (Werner-Heisenberg-Institut), M\"unchen, Germany}
\newcommand{\LLNL}{Lawrence Livermore National Laboratory, Livermore, CA 94550, USA}
\newcommand{\MPIS}{Max-Planck-Institut f\"{u}r Sonnensystemforschung, G\"{o}ttingen, Germany}
\newcommand{\Bogazici}{Bogazici University, Istanbul, Turkey.}

\newcommand{\XFEL}{European XFEL GmbH, Notkestrasse 85, 22607 Hamburg, Germany.}
\newcommand{\ECU}{Excellence Cluster Universe, Technische Universit\"{a}t M\"unchen, Garching, Germany.}

\newcommand{\BenGurion}{Physics Department, Ben-Gurion University of the Negev, Beer-Sheva 84105, Israel.}
\newcommand{\Korea}{School of Space Research, Kyung Hee University, Yongin, Republic of Korea.}
\newcommand{\ESS}{European Spallation Source ESS AB, Lund, Sweden.}

\newcommand{\Rijeka}{Physics Department and Center for Micro and Nano Sciences and Technologies, University of Rijeka, Radmile Matejcic 2, 51000 Rijeka, Croatia.}
\newcommand{\EPFL}{Laboratoire de Transfert de Chaleur et de Masse, \'Ecole Polytechnique F\'ed\'erale de Lausanne (EPFL), Lausanne, Switzerland.}
\newcommand{\California}{Dep. of Physics and Astronomy, University of California, Irvine, CA 92697, USA.}

\begin{document}

\widetext
\leftline{Prepared to be submitted in PRL}

\preprint{APS/123-QED}

\title{New solar axion search in CAST with $^4$He filling}

\author{    M.~Arik}\altaffiliation[Present addr.: ]{\Bogazici}\affiliation{\Dogus}
\author{    S.~Aune  }\affiliation{\Saclay}
\author{    K.~Barth  }\affiliation{\CERN}
\author{    A.~Belov  }\affiliation{\INR}
\author{    H.~Br\"auninger  }\affiliation{\MPE}
\author{    J.~Bremer  }\affiliation{\CERN}
\author{    V.~Burwitz  }\affiliation{\MPE}
\author{    G.~Cantatore  }\affiliation{\Trieste}
\author{    J.~M.~Carmona  }\affiliation{\Zaragoza}
\author{    S.~A.~Cetin  }\affiliation{\Dogus}
\author{    J.~I.~Collar  }\affiliation{\Chicago}
\author{    E.~Da~Riva  }\affiliation{\CERN}
\author{    T.~Dafni  }\affiliation{\Zaragoza}
\author{    M.~Davenport  }\affiliation{\CERN}
\author{    A.~Dermenev  }\affiliation{\INR}
\author{    C.~Eleftheriadis  }\affiliation{\Thessaloniki}
\author{    N.~Elias  }\altaffiliation[Present addr.: ]{\ESS}\affiliation{\CERN}
\author{    G.~Fanourakis  }\affiliation{\Demokritos}
\author{    E.~Ferrer-Ribas  }\affiliation{\Saclay}
\author{    J.~Gal\' an  }\affiliation{\Saclay}
\author{    J.~A.~Garc\' ia  }\email[Corresponding author: ]{jagarpas@unizar.es}\affiliation{\Zaragoza}
\author{    A.~Gardikiotis  }\affiliation{\Patras}
\author{    J.~G.~Garza  }\affiliation{\Zaragoza}
\author{    E.~N.~Gazis  }\affiliation{\Athens}
\author{    T.~Geralis  }\affiliation{\Demokritos}
\author{    E.~Georgiopoulou  }\affiliation{\Patras}
\author{    I.~Giomataris  }\affiliation{\Saclay}
\author{    S.~Gninenko  }\affiliation{\INR}
\author{    M.~G\' omez~Marzoa  }\altaffiliation[Also at.: ]{\EPFL}\affiliation{\CERN}
\author{    M.~D.~Hasinoff  }\affiliation{\Vancouver}
\author{    D.~H.~H.~Hoffmann  }\affiliation{\Darmstadt}
\author{    F.~J.~Iguaz  }\affiliation{\Zaragoza}
\author{    I.~G.~Irastorza  }\affiliation{\Zaragoza}
\author{    J.~Jacoby  }\affiliation{\Frankfurt}
\author{    K.~Jakov\v ci\' c  }\affiliation{\Zagreb}
\author{    M.~Karuza  }\affiliation{\Trieste}\affiliation{\Rijeka}
\author{    M.~Kavuk}\altaffiliation[Present addr.: ]{\Bogazici}\affiliation{\Dogus}
\author{    M.~Kr\v{c}mar  }\affiliation{\Zagreb}
\author{    M.~Kuster  }\altaffiliation[Present addr.: ]{\XFEL}\affiliation{\MPE}\affiliation{\Darmstadt}
\author{    B.~Laki\'{c}  }\affiliation{\Zagreb}
\author{    J.~M.~Laurent  }\affiliation{\CERN}
\author{    A.~Liolios  }\affiliation{\Thessaloniki}
\author{    A.~Ljubi\v{c}i\'{c}  }\affiliation{\Zagreb}
\author{    G.~Luz\'on  }\affiliation{\Zaragoza}
\author{    S.~Neff  }\affiliation{\Darmstadt}
\author{    T.~Niinikoski  }\altaffiliation[Present addr.: ]{\ECU}\affiliation{\CERN}
\author{    A.~Nordt  }\altaffiliation[Present addr.: ]{\ESS}\affiliation{\MPE}\affiliation{\Darmstadt}
\author{    I.~Ortega  }\affiliation{\Zaragoza}\affiliation{\CERN}
\author{    T.~Papaevangelou  }\affiliation{\Saclay}
\author{    M.~J.~Pivovaroff  }\affiliation{\LLNL}
\author{    G.~Raffelt  }\affiliation{\MPP}
\author{    A.~Rodr\'\i guez  }\affiliation{\Zaragoza}
\author{    M.~Rosu  }\affiliation{\Darmstadt}
\author{    J.~Ruz  }\affiliation{\LLNL}
\author{    I.~Savvidis  }\affiliation{\Thessaloniki}
\author{    I.~Shilon  }\altaffiliation[Also at.: ]{\BenGurion}\affiliation{\Zaragoza}\affiliation{\CERN}
\author{    S.~K.~Solanki  }\altaffiliation[Sec. Affiliation: ]{\Korea}\affiliation{\MPIS}
\author{    L.~Stewart  }\affiliation{\CERN}
\author{    A.~Tom\' as  }\affiliation{\Zaragoza}
\author{    T.~Vafeiadis  }\affiliation{\CERN}\affiliation{\Thessaloniki}\affiliation{\Patras}
\author{    J.~Villar  }\affiliation{\Zaragoza}
\author{    J.~K.~Vogel  }\affiliation{\LLNL}
\author{    S.~C.~Yildiz  }\altaffiliation[Present addr.: ]{\California}\affiliation{\Dogus}
\author{    K.~Zioutas  }\affiliation{\CERN}\affiliation{\Patras}

\collaboration{CAST Collaboration} 

\date{\today}

\begin{abstract}
The CERN Axion Solar Telescope (CAST) searches for $a\to\gamma$
conversion in the 9~T magnetic field of a refurbished LHC test magnet
that can be directed toward the Sun. Two parallel magnet bores can be filled with
helium of adjustable pressure to match the X-ray refractive mass
$m_\gamma$ to the axion search mass $m_a$. After the vacuum phase
(2003--2004), which is optimal for $m_a\lesssim0.02$~eV, we used $^4$He in
2005--2007 to cover the mass range of 0.02--0.39~eV and $^3$He in 2009--2011 to scan from
0.39--1.17~eV. After improving the detectors and shielding, we
returned to $^4$He in 2012 to investigate a narrow $m_a$ range around
0.2~eV (``candidate setting'' of our earlier search) and
0.39--0.42~eV, the upper axion mass range reachable with $^4$He, to ``cross
the axion line'' for the KSVZ model. We have improved the limit on the
axion-photon coupling to $g_{a\gamma}< 1.47\times10^{-10}~{\rm
  GeV}^{-1}$ 
(95\% C.L.), depending on the pressure
settings. Since 2013, we have returned to vacuum and aim for a
significant increase in sensitivity.
\end{abstract}


\maketitle

{\em Introduction}---The low-energy frontier of elementary particle
physics \cite{Jaeckel:2010ni, Ringwald:2012hr, Hewett:2012ns, Baker:2013zta, Essig:2013lka, Agashe:2014kda}
includes numerous experimental efforts, ranging
from the search for neutrino-less double beta decay and searches for
nucleon and electron electric dipole moments all the way to searches
for new low-mass bosons. The best-motivated case for the latter
remains the axion, the pseudo Nambu-Goldstone boson of a new broken global
$U(1)$ symmetry, that is required in the context of the Peccei-Quinn
mechanism to explain why CP-violating effects are extremely small or
absent in QCD \cite{Peccei:2006as, Kim:2008hd}. The axion is also an
excellent candidate for the cold dark matter of the universe
\cite{Sikivie:2006ni} and ongoing experimental searches as well as new
efforts have recently gained fresh momentum \cite{Asztalos:2009yp, Asztalos:2011bm, vanBibber:2013ssa,
  Rybka:2014xca, Cho:2013dwa, Baker:2011na, Horns:2012jf,
  Jaeckel:2013sqa, Jaeckel:2013eha, Horns:2013ira, Sikivie:2013laa, Rybka:2014cya,CAPP,SPSCCAST2014}.
On the other hand, the phenomenology of axions has inspired the proposition
of axion-like particles (ALPs) and other WISPs (weakly interacting
sub-eV particles) with additional theoretical motivations from string
theory and cosmology. Experimental searches for low-mass
bosons include precision searches for new long-range forces
\cite{Moody:1984ba, Raffelt:2012sp, Adelberger:2013faa, Leslie:2014mua} 
and oscillating nucleon electric dipole
moments~\cite{Graham:2011qk, Graham:2013gfa, Budker:2013hfa}.

\begin{figure}
\includegraphics[width=1.\columnwidth]{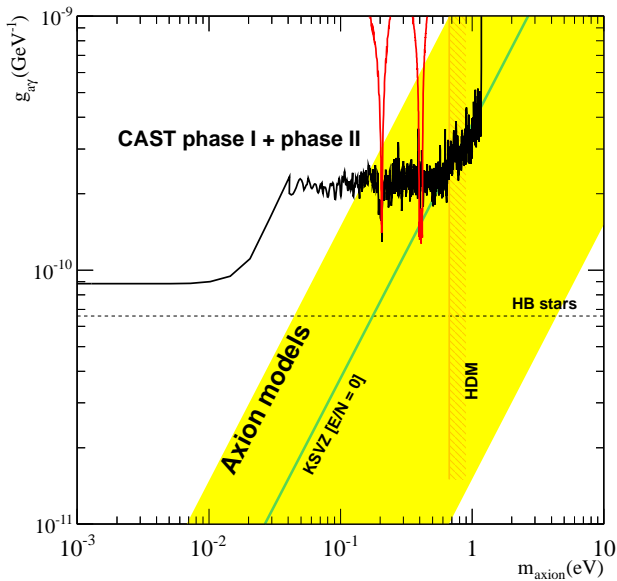}
\caption{Exclusion regions in the $m_a$--$g_{a\gamma}$--plane achieved
  by CAST in the vacuum phase~\cite{Zioutas:2004hi,Andriamonje:2007ew}
  and with $^4$He \cite{Arik:2008mq} and $^3$He
  \cite{Arik:2011rx,Arik:2013nya} filling.  We also show constraints
  from horizontal branch (HB) stars~\cite{Raffelt:2006cw,Ayala:2014pea} 
  (a similar limit stems from blue-loop suppression in massive
  stars~\cite{Friedland:2012hj}), and the hot dark matter (HDM)
  bound~\cite{Hannestad:2005df,Hannestad:2007dd,Melchiorri:2007cd,Hannestad:2008js,Hannestad:2010yi,Archidiacono:2013cha}. The yellow band represents
  typical theoretical models with $\left|E/N-1.95\right|=0.07$--7. The
  green solid line corresponds to $E/N=0$ as in the KSVZ model
  \cite{Kim:1979if, Shifman:1979if}, a typical hadronic axion model.
  In red we show our new limits near $m_a=0.2$ and 0.4~eV from our
  2012 data taking campaign with $^4$He gas.}\label{fig:limits}
\end{figure}

The most promising technique to search for axions to date remains Sikivie's idea to use the
two-photon vertex of axions or ALPs and look for their conversion to
photons in a strong external magnetic field $B$ (also referred to as the Primakoff effect) \cite{Sikivie:1983ip}. The
smallness of the coupling strength $g_{a\gamma}$ can be overcome by
coherent conversion in a macroscopic volume. For more than ten years,
the CERN Axion Solar Telescope (CAST) has pursued this idea in the
form of a large axion helioscope \cite{Zioutas:2004hi,
  Andriamonje:2007ew, Arik:2008mq, Arik:2011rx, Arik:2013nya}.
Earlier helioscope searches were conducted in Brookhaven
\cite{Lazarus:1992ry} and Tokyo \cite{Moriyama:1998kd, Inoue:2002qy,
  Inoue:2008zp}. In CAST, a refurbished LHC dipole magnet (length
9.26~m, field strength 9.0~T) is mounted to follow the Sun for
approximately 1.5~h at both sunrise and sunset. In this way various
X-ray detectors at both ends of the magnet are used to look at the Sun.

For a fixed axion-photon coupling strength $g_{a\gamma}$, the
$a$-$\gamma$ conversion probability decreases when the $a$-$\gamma$
oscillation length becomes smaller than the magnet length, which limits
CAST's sensitivity to an axion mass range $m_a \lesssim 0.02$~eV. For larger $m_a$ values, the
transition rate can be restored by providing photons with a refractive
mass using a low-$Z$ gas filling~\cite{vanBibber:1988ge}. The gas
density (i.e.\ pressure) defines the search mass and one needs to step
through many pressure settings to search a broad $m_a$ range.

CAST has taken data since 2003 and produced the exclusion plot
shown in Fig.~\ref{fig:limits}. The vacuum phase (2003--2004)
achieved the best sensitivity up to $m_a \sim 0.02$~eV because of the long exposure
time. With $^4$He filling (2005--2007), the maximum search mass was extended
to $\sim0.4$~eV, corresponding to the $^4$He vapor pressure at cryogenic
temperatures. With $^3$He in 2009--2011 we finally reached
1.17~eV, which exceeds the cosmological hot dark matter limit for
axions \cite{Hannestad:2005df,Hannestad:2007dd,Melchiorri:2007cd,Hannestad:2008js,Hannestad:2010yi,Archidiacono:2013cha}. 
For $m_a>0.6$~eV our $^3$He search has reached the ``axion
line'' for a typical hadronic axion model \cite{Kim:1979if, Shifman:1979if} (E/N =0, see Fig.~\ref{fig:limits}),
i.e., the locus of $g_{a\gamma}$ vs.\ $m_a$ which is motivated by QCD
axions, as opposed to more general ALPs.

To make the experimental progress pushing the sensitivity to lower coupling constants of
Fig.~\ref{fig:limits}, a much larger helioscope, the International
Axion Observatory (IAXO) \cite{Irastorza:2011gs, Shilon:2012te, Armengaud:2014gea}, 
has been envisioned and, for very small $m_a$, a
large-scale photon-regeneration experiment ALPS-II
\cite{Bahre:2013ywa} has been put forward. Until these next-generation projects
become operational, CAST can push its boundaries by reducing
background levels and increasing the exposure time. Therefore, CAST
has engaged in a programme of technical improvements and renewed data
taking to achieve these goals.
 
We report here on results from our 2012 data taking campaign
with $^4$He gas. At each pressure setting, we integrated for
about 7.5~h per detector (5 solar trackings), in
contrast to 1.5~h (1 solar tracking) in our earlier $^4$He campaign. 
On the sunset side,
new Micromegas detectors with improved shielding and veto were
installed, decreasing the background level by an approximate factor
of~4 \cite{Aune:2011}. The search masses were selected to cover a previous
``candidate setting'' ($m_a\sim 0.2$~eV), where unusually many events
above background had appeared, and the upper end of what can be reached
with $^4$He ($m_a\sim 0.4$~eV) in order to reach the KSVZ (Kim, Shifman, Vainshtein, Zakharov model) axion line.


{\em System description}---After completing the $^3$He phase, the gas was removed from the system. The magnet bores were then filled with  $^4$He to take advantage of the sophisticated gas metering
system capable of filling the cold bore in small steps with a reproducibility of better than 100~ppm \cite{Arik:2011rx}.

As explained in Ref.~\cite{Arik:2013nya}, at increasingly higher gas densities 
(14--108~mbar at 1.8~K), $^3$He gas dynamics (convection and buoyancy) at the ends of the 
cold bore affect the density distribution along the cold bore. This
effect progressively shortens 
the region with uniform density (effective coherence length) and also causes variations in the 
central density during the tilting of the magnet. In the $^4$He case, the densities involved
are relatively low (up to 15.5~mbar at 1.8~K) and these effects are negligible.

For $^4$He, the central gas density was calculated from the equation of state
 of the $^4$He gas using the measured cold bore temperature, gas pressure and magnet vertical
angle. The coherence length was taken as the magnet length of 9.26~m.

Whilst Computational Fluid Dynamics (CFD) simulations have not been performed for $^4$He, the $^3$He 
CFD simulations at a similar pressure over various vertical angles gave effective coherence lengths
above $L_{\rm eff}=8$~m. The main source of systematic error in the present result, according to our previous studies~\cite{Andriamonje:2007ew,Arik:2013nya,Arik:2008mq} is given by the uncertainties on the gas dynamics inside the magnet bores. In order to estimate the effect of such uncertainties the upper limit calculation has been redone using a reduced effective coherence length $L_{\rm eff}=8$~m for all vertical angles. This value represents a worst case scenario of the effect of gas dynamics on $L_{\rm eff}$ according to CFD simulations performed in Ref.~\cite{Arik:2013nya}.

The X-ray detectors installed at CAST during the 2012 data taking
campaign were three Micromegas detectors of the microbulk type
\cite{Abbon:2007,Aune:2009,Andriamonje:2010,Galan:2010}
(one in the sunrise and two in the sunset side) and a pn-CCD
detector in the focal plane of an X-ray telescope \cite{Kuster:2007}
on the sunrise side. While the detectors on the sunrise side 
remained unchanged since the previous data taking campaign, the
Micromegas detectors on the sunset side were upgraded, improving
the background levels of the detectors. This is the result of low background techniques developed for the Micromegas detectors \cite{Aune:2011}, where different strategies were exploited: the manufacturing technology of the novel microbulk Micromegas, the intrinsic radiopurity of the detectors \cite{Cebrian:2011}, the discrimination algorithms on the analysis, and different shielding strategies.

The upgrade focuses on reducing the contribution of the
environmental gamma flux. The lead shielding thickness was increased 
(from 25~mm to 100~mm) and the design is more compact, improving the shielding
around the pipes to the magnet. The inner copper shielding was
increased from 5~mm to 10~mm of copper. It is connected to the magnet
bores by a 10~mm thick copper pipe, which has an inner
polytetrafluoroethylene coating of 2.5~mm thickness in order to
attenuate the 8~keV copper fluorescence. In addition, the aluminum
strongback was replaced by a more radiopure copper one, and all the
components close to the detector have been carefully selected and
cleaned. A plastic scintillator was installed on the top
of the shielding, allowing the discrimination of background events induced by cosmic muons in the detectors (see Ref.~\cite{Aune:2011} for more details).

\begin{figure}
\includegraphics[width=1.\columnwidth]{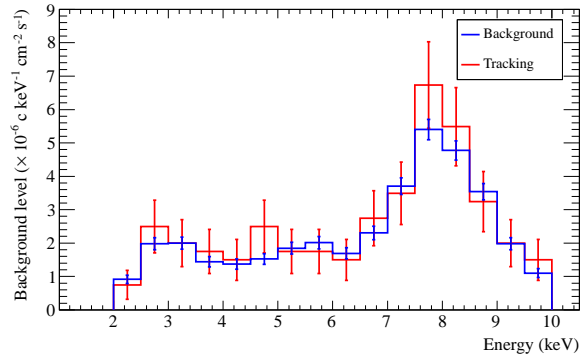}
\caption{ Comparison between background (blue bars) and tracking (red bars) spectra of one Micromegas detectors installed in the sunset side.}\label{fig:SS2Spectra}
\end{figure}

These upgrades reduce the background level to $1.3$ and $1.7\times
10^{-6}$ c cm$^{-2}$ keV$^{-1}$ s$^{-1}$ for the sunset detectors 1
and~2, respectively, i.e., about a factor 4 lower than before,
improving the signal-to-noise ratio by a factor of~2. Background and tracking spectra of one of the detectors in the sunset side are shown in Fig.~\ref{fig:SS2Spectra}. As seen both levels are compatible within their error. The background is dominated by the copper fluorescence at 8 keV, its escape peak above 5 keV and the Argon fluorescence around 3 keV. According to our current understanding of the background \cite{Aune:2011,Garcia:2013,Tomas:2013phd}, these fluorescences are induced by secondary particles generated in the inner materials by external radiation, mainly muons that are not tagged by the active veto system. On the other hand, the detector response has been fully characterized by a detailed simulation \cite{Tomas:2013phd} and calibrations at different energies in an X-ray beam \cite{Garza:2013}. Moreover, daily calibrations allow us to monitor the performance of the detectors during the data taking period (see Ref. \cite{Aune:2011} for more details).

The extensive research program for the background reduction continues; different measurements in special set-ups have been performed, such as the ones underground at the Laboratorio Subterr\'aneo de Canfranc \cite{Tomas:2012}, where the cosmic muons are highly suppressed, and the test at surface level~\cite{Garza:2013}. These special test benches, together with the simulations \cite{Tomas:2013phd}, were crucial in order to understand the origin of the background in the Micromegas detectors. Additional upgrades have been implemented at CAST for the re-scanned vacuum phase that started in 2013, like the AFTER \cite{Calvet:2008} front end electronics and a new design of the active shielding on the sunset side, with improved background levels on the Micromegas detectors further.


{\em Data analysis and results.}---The 
results presented in this work are related to the data acquired with three Micromegas detectors
during 2012 using $^4$He as a buffer gas. The total exposure time in axion-sensitive conditions
 (solar tracking) was about 147 h per detector with a background time of 2277~h per detector.
  In the first part of the 2012 data taking campaign, the axion mass range $0.39 <m_a < 0.42$~eV was scanned,
which corresponds to the pressure range 13.9--15.5~mbar at 1.8~K. 
During this period, 17 pressure settings were
covered with a step size of 0.1~mbar and an effective average exposure time of $\sim$24~h per setting. In the second part
of the data taking campaign, a single setting at $m_a \simeq 0.20$~eV was covered during $\sim30$~h.

The analysis of the data was performed in the same manner as in our previous
publications \cite{Arik:2011rx, Arik:2013nya}. In order to take into
account the variations of the density inside the magnet bores during the trackings, an unbinned likelihood was implemented, where the likelihood function can be expressed as
\begin{equation}
\log{\mathcal{L}} \propto -R_T + \sum_{i}^n \log{ R(t_i,E_i,d_i)}\,.
\end{equation}
Here, $R_T$ is the expected number of counts from the axion to photon conversion over all the exposure time, energy and detectors. The sum is over each of the $n$ detected counts during the tracking time, for an expected rate $R(t_i,E_i,d_i)$ as a function of the event time $t_i$, energy $E_i$ and detector $d_i$, given by the expression
\begin{equation}
R(t,E,d) = B_d + S(t,E,d)\,,
\end{equation}
where $B_d$ is the background level of the detector $d$. $S(t,E,d)$ is the expected rate from axion conversion in the detector $d$ given by
\begin{equation}
S(t,E,d) = \frac{d \Phi_{a}} {d E} P_{a\rightarrow\gamma} \epsilon_d\,.
\end{equation}
Here, $\epsilon_d$ is the detector efficiency, $\frac{d \Phi_{a}}{d E}$ is the differential solar axion flux, which can be parametrized \cite{Raffelt:2006cw} by the expression 
\begin{equation}
\frac {d \Phi_{a}} {d E} = 6.02 \times 10^{10} g_{10}^{2} \frac{E^{2.481}} {e^{E/1.205}} \left [ \hbox{cm}^{-2} \hbox{s}^{-1} \hbox{keV}^{-1} \right ]
\end{equation}
with $g_{10}=g_{a\gamma}/(10^{-10}\mbox{GeV}^{-1})$ and energies in keV. $P_{a\rightarrow\gamma}$ is the axion to photon conversion probability inside a strong magnetic field \cite{vanBibber:1988ge}, given by
\begin{equation}\label{eq:ConvProb}
P_{a\rightarrow\gamma} = \left (\frac {g_{a\gamma} B}{2} \right)^2 \frac{ 1 + e^{-\Gamma L} - 2 e^{-\Gamma L/2} \cos(qL)}{q^2 + \Gamma^2/4}
\end{equation}
where $q = |m_a^2 - m_\gamma^2|/(2E)$ is the axion-photon momentum transfer in a magnetic field $B$ and $\Gamma$ is the absorption coefficient in the buffer gas.

As explained in Ref.~\cite{Arik:2008mq}, the dependence on $m_a$ in the expressions above is included in the conversion probability given in Eq.~(\ref{eq:ConvProb}), where $P_{a\rightarrow\gamma}$ is enhanced for axion masses that match the refractive photon mass $m_\gamma$ as determined by the buffer gas density. Therefore for a given axion mass $m_a$, only the counts for which the coherence condition is fulfilled will contribute to the likelihood function $\log{\mathcal{L}}$. 

\begin{figure}
\includegraphics[width=1.\columnwidth]{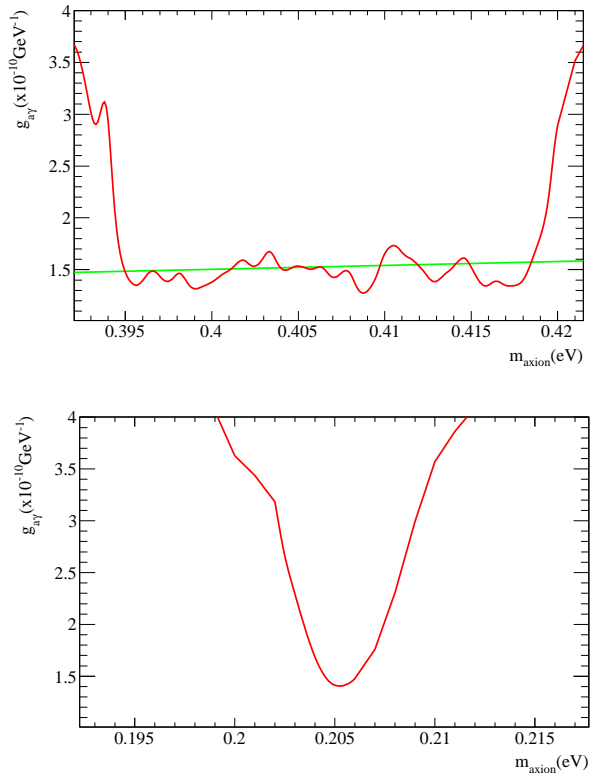}
\caption{Expanded view of the limit achieved during the 2012 CAST data taking campaign, using $^4$He as buffer gas. The plot on the bottom corresponds to the ``candidate setting'' at 0.2~eV, while the plot on the top is the excluded region above 0.4~eV. The green line represents the KSVZ benchmark model with $E/N=0$.}\label{fig:ExcplotZoom}
\end{figure}

By maximizing $\log{\mathcal{L}}$ a best-fit value $g_{\rm min}^4$ is
obtained. This value is compatible with the absence of a signal in the entire axion mass range
and thus an upper limit on $g_{a\gamma}$ is extracted, by
integrating the Bayesian posterior probability from zero up to 95\%
with a flat prior in $g_{a\gamma}^{4}$. The computed upper limit for
several values of $m_a$ is displayed in red in Fig. \ref{fig:limits}. A close
view of the excluded region is shown in Fig. \ref{fig:ExcplotZoom},
where only the axion mass range scanned during the 2012 data taking
campaign is included.

This demonstrates that CAST improves its previous limit for axion masses
$0.39<m_a<0.42$~eV, down to an average value of the coupling constant
$g_{a\gamma}\sim1.47 \times 10^{-10}$~GeV$^{-1}$. Moreover, the ``candidate setting'' around 0.2~eV was rescanned and in the absence of an axion signature a coupling limit of $g_{a\gamma} < 1.40 \times 10^{-10}$~GeV$^{-1}$ at a 95$\%$ of C.L. was derived. In contrast to the previous results~\cite{Arik:2008mq,Arik:2011rx,Arik:2013nya}, the structure of the excluded region has a smooth shape, because of considerably larger exposure time per step, resulting in smaller statistical fluctuations.


{\em Conclusions.}---After finishing its initial mission of searching
for solar axions up to the largest $m_a$ value that could be
reasonably reached with $^3$He, CAST has embarked on a programme of
technical improvements aimed at increasing its sensitivity. Here we
 reported the first results from a new $^4$He run with significantly
reduced background rates in two narrow $m_a$ bands around 0.2 and
0.4~eV. The bounds on $g_{a\gamma}$ were significantly improved in these regions,
allowing us to cross the KSVZ axion line near the highest possible
$^4$He pressure.

Since 2013, we have returned to vacuum measurements, i.e., the low-mass
regime, $m_a<0.02$~eV. With various further improvements, notably an
 additional X-ray telescope, we aim at a sensitivity for $g_{a\gamma}\simeq0.6\times10^{-10}~{\rm GeV}^{-1}$ for these low masses that could even surpass a recently
improved stellar-evolution bound from the helium-burning lifetime
of globular-cluster stars~\cite{Ayala:2014pea}. This low-mass regime, while far away
from the ``axion line,'' is important in the context of ALPs, e.g. the propagation of TeV
gamma-rays in large-scale astrophysical magnetic fields can be addressed
by a possible photon-ALP conversion.

{\em Acknowledgments.}---We thank CERN for hosting CAST and for
technical support to operate the magnet and cryogenics. We thank the
CERN CFD team for their essential contribution to the CFD work. We
acknowledge support from NSERC (Canada), MSES (Croatia), 
CEA (France), BMBF (Germany) under the
grant numbers 05 CC2EEA/9 and 05 CC1RD1/0 and DFG (Germany) under
grant numbers HO 1400/7-1 and EXC-153, GSRT (Greece), NSRF:
Heracleitus II, RFFR (Russia), the Spanish Ministry of Economy and
Competitiveness (MINECO) under Grants No.\ FPA2008-03456 and
No.\ FPA2011-24058. This work was partially funded by the European
Regional Development Fund (ERDF/FEDER), the European Research Council
(ERC) under grant ERC-2009-StG-240054 (T-REX), Turkish Atomic Energy
Authority (TAEK), NSF (USA) under Award No.\ 0239812 
and NASA under the grant number NAG5-10842. Part of this
work was performed under the auspices of the U.S.\ Department of
Energy by Lawrence Livermore National Laboratory under Contract
No.\ DE-AC52-07NA27344.

\end{document}